\documentclass[english]{llncs}
\usepackage[T1]{fontenc}
\usepackage[latin9]{inputenc}
\usepackage{geometry}
\usepackage{graphicx}
\usepackage{framed}
\usepackage{hyperref}
\usepackage{subfig}
\usepackage[linesnumbered]{algorithm2e}
\usepackage{comment}
\usepackage{pgfplots}
\usepackage{tikz}
\usepackage{xintexpr}
\usepackage[normalem]{ulem}
\newcommand{\kaoutar}[1]{{\color{magenta}{\bf kaoutar:} #1}}

\usepackage{babel}
\begin{document}
\title{Arma: Byzantine Fault Tolerant consensus with Linear Scalability}
\author{Yacov Manevich}
\institute{IBM Research, Zurich}
\maketitle
\begin{abstract}
Arma\footnote{Means ``Chariot'' in Greek. Similarly to a chariot that can tow
more when horses are added, Arma totally orders more user transactions
the more hardware it is added.} is a Byzantine Fault Tolerant (BFT) consensus system designed to
achieve linear scalability across all hardware resources: network
bandwidth, CPU, and disk I/O. As opposed to preceding
BFT protocols, Arma separates the dissemination and validation of
client transactions from the consensus process, restricting the latter to totally ordering only metadata of batches of transactions. This separation enables
each party to distribute compute and storage resources for transaction validation, dissemination and disk I/O among multiple machines, resulting in linear scalability.
Additionally, Arma ensures censorship resistance by imposing a maximum
time limit for the inclusion of client transactions. We build a prototype implementation of Arma and evaluate its performance experimentally. Our results show that Arma totally orders over 100,000 transactions per second when deployed in a WAN setting and integrated into Hyperledger Fabric.

\end{abstract}

\section{Introduction}
%\textcolor{blue}{lli: I would recommend something more compact but to the point… Something with same structure/different wording maybe with the text below?}

Reaching consensus in the presence of Byzantine failures is a problem introduced by Lamport et al.~\cite{ByzGen} and has since been long investigated by the scientific community. Byzantine Fault Tolerant (BFT) consensus has gained more attention with the advent of Distributed Ledger Technologies and the extended use of the latter for financial asset exchange applications, i.e., scenarios that desire the transparency and high degree of resilience that BFT consensus offers. However, initial works of BFT protocols suffered from poor performance, especially with a high number of participants. 
In recent years, substantial efforts in both academic and industrial research have demonstrated that BFT systems are capable of achieving significant throughput at large scale.

%\subsubsection{Non monolithic consensus}

A prevalent characteristic among consensus protocols is their monolithic
architecture, where each participating party consolidates all the
subroutines of the protocol within a single node. Consequently, every
party operates a node that encompasses all aspects of the protocol.
Several reasons can account for this approach. Firstly, implementing
and analyzing a distributed consensus protocol involving multiple
parties is inherently challenging, and having each party manage multiple
nodes further exacerbates the complexity. Secondly, many researchers
gauge protocol efficiency based on the overall number of messages
sent during the protocol or by each party, often overlooking the significant
discrepancy in message sizes and their transmission efficiency.
Indeed, the network bandwidth a protocol requires from each node
is often the limiting factor for the throughput and not the number of messages sent.
Moreover, most consensus protocols are evolutions of earlier protocols
and can be traced back to seminal works such as PBFT \cite{PBFT} or Paxos \cite{Lamport2001PaxosMS}, which
possess a monolithic structure.

While the monolithic architecture is relatively straightforward to
implement and analyze, it inherently constraints the performance of
the system. Eventually, the machine of each party will reach its limit in
terms of CPU and storage operations per second.
Only recently, researchers have proposed a consensus protocol \cite{Narwhale} that
allows each party to distribute its execution across multiple machines,
where not all nodes are created equal. This design enables the horizontal
scaling of all resources required for consensus, including CPU, storage
I/O rate, and network bandwidth. In such systems, each party operates
several nodes, and the various subroutines of the consensus protocol
are executed on nodes based on their specific roles. However, current implementations 
of such systems do not implement censorship resistance, and don't prevent transaction de-duplication.

\subsubsection{The contribution of this work}

This paper introduces Arma, a novel consensus protocol that builds on
ideas from the recent work of Danezis, Lefteris, Sonnino and Spiegelman \cite{Narwhale} to achieve linear
scalability. Distinguishing itself from the previous work, Arma incorporates
censorship resistance and transaction de-duplication mechanisms. Once a client
submits a transaction to Arma, it can be assured of its finalization.
As a result, Arma not only demonstrates high throughput but also enhances
the user experience, by eliminating the need for client applications
to include logic for tracking transaction finalization. This advancement
opens up possibilities for constructing payment systems where correct
clients can effortlessly initiate transactions without the necessity
of monitoring their success in real-time.

\subsubsection{Brief overview of the Arma protocol}

In the Arma consensus protocol, transaction dissemination and validation are entrusted to separate groups of nodes known as shards. Each shard includes a representative from every party and exclusively manages a specific subset of the transaction hash space.

The nodes within these shards provide attestations concerning persisted batches of transactions or votes related to node misbehavior. These attestations and votes are then submitted into a Byzantine Fault Tolerant (BFT) consensus protocol. By totally ordering these attestations and votes, Arma nodes gain collective knowledge regarding which batches of transactions have been safely persisted by a sufficient number of nodes and identifies nodes displaying faulty behavior, such as deliberate transaction censorship or being unreachable.
With the attestations and votes fully ordered through the BFT consensus protocol, an ordering between batches in different shards is achieved, subsequently leading to an implicit total order between all transactions.\\

%\subsubsection{Paper outline}

\noindent \textbf{Paper Outline.}The subsequent section of this paper delves into the evolution of
Byzantine Fault Tolerant (BFT) consensus protocols. Section 3 provides
a comprehensive overview of the design of the Arma protocol, highlighting
its key components and mechanisms. Following that, in section 4, it is explained how a prototype which integrates Arma into the Hyperledger Fabric \cite{HLF} ordering service was made, and discusses its performance evaluation. In section 5 we give a formal analysis of Arma's properties, and in Section 6 we conclude and discuss future work.

\section{Related work}

This section discusses prior work, which is relevant not
only to comprehend the current state of the art in high-performance
Byzantine Fault Tolerant (BFT) systems but also to understand the design decisions underpinning the Arma system.

\subsection{Practical Byzantine Fault Tolerance}

In contrast to permissionless blockchains like Bitcoin and Ethereum,
systems designed for retail and enterprise use cases often possess
a well-defined membership structure and impose stringent throughput
requirements. Among the various Byzantine Fault Tolerant (BFT) protocols,
the PBFT \cite{PBFT} protocol stands out as a prominent choice for such settings,
serving as the foundation upon which subsequent works build.
A key characteristic shared by these protocols is the establishment
of a total order by having a designated leader node broadcast a batch
of client transactions. However, this inherent asymmetry among the
nodes limits the protocol's throughput, as it becomes constrained
by the network bandwidth of the leader node.

\begin{figure}[h]%
	\begin{framed}
		\centering
		\subfloat[\centering Each party (a,b,c,d) runs a single machine with co-located BFT instances]{{\includegraphics[width=5cm]{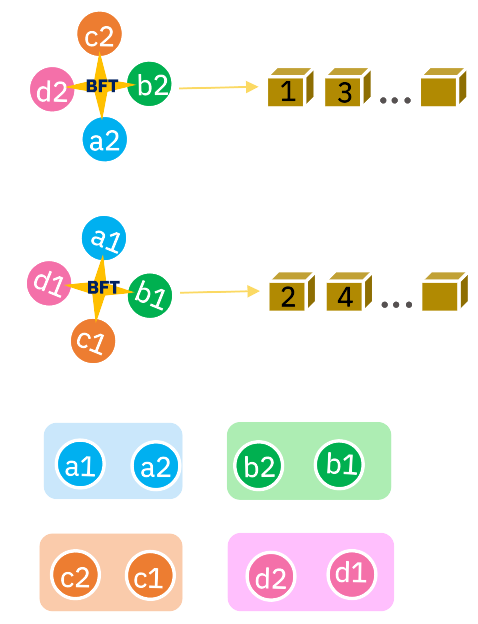} }}%
		\qquad
		\qquad
		\qquad
		\subfloat[\centering Each party (a,b,c,d) runs several machines, each executing a dedicated protocol instance]{{\includegraphics[width=7cm]{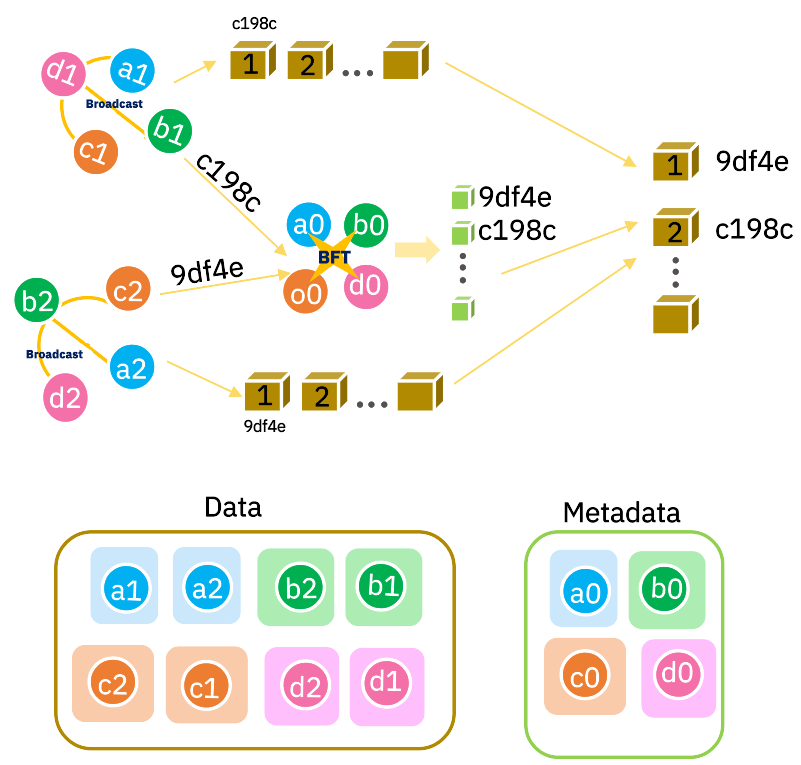} }}%
		\caption{(a) Multi instance BFT  vs (b) Distributed Multi Instance BFT }%
		\label{fig:multi}
	\end{framed}
\end{figure}

\subsection{Multi Instance BFT}

In their pursuit of a Byzantine Fault Tolerant system that aligns
with enterprise use cases, researchers have consistently sought
to identify bottlenecks in existing protocols and propose methods
to mitigate them. A notable example of significant performance enhancement
can be found in the techniques employed by Mir-BFT \cite{Mir-BFT}. 

Mir-BFT achieves
this by running multiple parallel instances of PBFT \cite{PBFT}, where each batch
sequence $i$ is specifically reserved for the PBFT instance $i\,\%\,k$,
with $k$ being the number of parallel PBFT instances. This approach
effectively induces a total order while addressing the identified
bottlenecks, resulting in notable performance improvements.

\subsubsection{Transaction de-duplication via hashing}

In Mir-BFT, every client transaction is assigned to a specific PBFT
instance by hashing it to a value within the range of $\left\{ 1,...,k\right\} $.
By running multiple PBFT instances, Mir-BFT effectively distributes
the network load that would typically be concentrated on a single
leader across multiple leaders. This distribution helps alleviate
the network bandwidth bottleneck associated with having a single leader.
Furthermore, the deterministic hashing of transactions to PBFT instances
ensures that duplicate transactions are prevented. This feature is
crucial not only for preventing denial-of-service attacks but also
for avoiding unnecessary duplication of total orderings for the same
transaction.

\subsubsection{Susceptibility to node crashes}

Running BFT instances in parallel can indeed increase throughput,
but it also introduces a higher vulnerability to node crashes. Let's
consider a protocol like Mir-BFT, where there are $k=\frac{n}{2}$
parallel instances of BFT, with n being the total number of nodes.
In this scenario, batch $b_{i}$ is assembled and broadcast by the
leader of instance $i\,\%\,k$ , and similarly, batch $b_{i+1}$ is
broadcast by the leader of instance $(i+1)\,\%\,k$. However, if the
leader of node $i$ crashes before broadcasting batch $b_{i}$, batch
$b_{i+1}$ cannot be delivered until a new leader is established in
instance $i\,\%\,k$. 

As a result, in such a protocol with k parallel PBFT instances, the
throughput drops to zero when any of the leaders among the $k$ different
instances crashes. In comparison, in a standard PBFT protocol, the
throughput drops to zero only when the leader node crashes. The increased susceptibility
to crashes in Mir-BFT's approach can have a significant impact on
the system's uptime availability.

\subsubsection{Throughput limitation causes}

In addition to transmitting batches over the network, nodes participating
in the consensus process perform other essential tasks. These tasks
include writing the batches to disk for crash fault tolerance, parsing
the client transactions within the batches, and verifying their integrity.
The speed at which batches are written to disk is influenced by the
underlying storage system of the node. On the other hand, transaction
verification is a computationally intensive task that relies on the
CPU's processing power. Both disk writing and transaction verification
play crucial roles in the overall performance of the consensus process.
Consequently, a protocol that is efficient in load balancing network bandwidth
across all parties may have its limiting factor shift towards CPU and disk I/O.

\subsection{Distributed Multi Instance BFT}

Running PBFT in parallel enhances horizontal scalability for the network
aspect, but it also introduces certain limitations. In this setup,
each node participates in multiple PBFT protocols, resulting in the
need to verify all transactions and write all batches to its underlying
storage. Consequently, the scalability of Mir-BFT is primarily limited
to the vertical scaling of CPU and storage I/O. Simply adding more
machines to the system cannot effectively address potential CPU or
storage I/O bottlenecks that may arise.

The works of Danezis, Lefteris, Sonnino and Spiegelman called ``Narwhal and Tusk''\cite{Narwhale}, however, offer insights into spreading
the load of storage I/O and CPU across multiple machines within the
same party, thereby overcoming these limitations. The key concept
is to distribute the tasks of transaction dissemination and validation
across multiple machines and subsequently establish a total order
for the corresponding metadata. By effectively separating these
responsibilities and achieving a distributed consensus on the metadata,
the aforementioned work \cite{Narwhale} enables the scalability of storage I/O and CPU by
leveraging the collective resources of multiple machines within a
party.

 Figure \ref{fig:multi} shows the architectural difference to the previous architecture, the Multi instance BFT.

\subsubsection{Totally ordering batch availability}

The actual dissemination of transactions is not conducted through
a full consensus protocol but rather through a sub-protocol known
as reliable broadcast. In \cite{Narwhale}, once a batch is persisted by $2F+1$ nodes
(where $F$ represents the maximum number of Byzantine faults the
system can tolerate), a certificate of availability is formed. This
certificate includes $2F+1$ signatures on the metadata of the batch.
The certificate of availability is then subjected to a real consensus
protocol to establish a total order. The total order of transactions
is derived from the order of the metadata of the batches.

Similar to a multi-instance Byzantine Fault Tolerant (BFT) system,
there can be multiple instances of the sub-protocol responsible for
transaction dissemination. Each instance can execute on its own machine,
and instances operated by the same party are not colocated on the
same machine. This distributed execution allows for scalability and 
better resource utilization across multiple machines, contributing to the
overall efficiency and performance of the consensus protocol.

\subsubsection{Crash fault resilience}

Unlike a multi-instance BFT approach, a distributed multi-instance
BFT design is less vulnerable to the crash of a party. In the event
of a node responsible for broadcasting batches experiencing a crash,
other instances can still completely order certificates of availability.
This resilience is attributed to the fact that the order among instances
transmitting transactions is determined by the consensus protocol
itself, rather than relying on a predetermined sequence allocation
as seen in multi-instance BFT systems like Mir-BFT.

One significant contribution of \cite{Narwhale} is the integration
of a fully asynchronous consensus protocol that operates without a
designated leader. This leaderless consensus protocol ensures that
transactions can still be finalized, regardless of which node in the
system experiences a crash. This feature enhances the fault tolerance
and availability of the system, as it allows for continuous progress
even in the presence of node failures.

\subsubsection{Lack of censorship resistance and de-duplication}

The approach  of \cite{Narwhale} does not inherently possess censorship
resistance or transaction de-duplication mechanisms. A malicious node
responsible for broadcasting batches can selectively ignore transactions
or intentionally slow down its execution, potentially leading to delays.
In such cases, the system relies on clients to resubmit their transactions
if they are not finalized within a designated timeout period. Furthermore,
malicious clients can submit their transactions to multiple instances,
resulting in wasteful resource utilization and potential duplication
of finalized transactions.

In contrast, the Arma system, which will be detailed in the next section,
builds upon the fundamental idea of \cite{Narwhale} and thus benefits from its inherent performance advantages, but also incorporates
mechanisms to achieve censorship resistance and transaction de-duplication.
By following the Arma protocol, clients can be certain that their
transactions will always be ordered, eliminating the need for timeout-based
resubmission.

\section{Architectural overview}
\label{sec:arch}

\begin{figure}[h]
	\begin{framed}
		\centering{{\includegraphics[width=12cm]{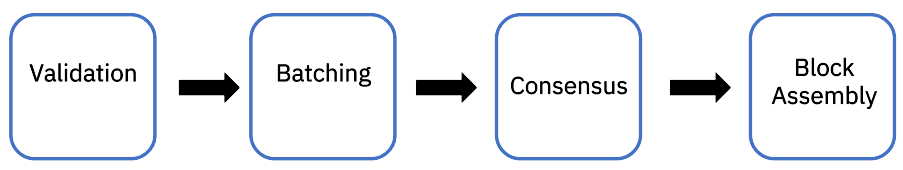} }}%
		\caption{The four stages of the Arma consensus protocol}%
		\label{fig:stages}
	\end{framed}
\end{figure}

Figure \ref{fig:stages} depicts the Arma protocol, that is broken down into four phases. In the first phase, transactions are \emph{validated} against some pre-defined, static system rules (e.g., this phase may include checks on the transaction format, and client signatures); after the validation phase completes, transactions are dispatched to nodes responsible for the \emph{Batching} phase, called \emph{batchers}. For scalability purposes, the transaction space is sharded, and there is a dedicated set of batchers for each shard. During batching, transactions are bundled into batches and persisted on disk. The batchers then create \emph{batch attestation shares} which are signatures over batch digests, and submit the batch attestation shares to the \emph{consensus} phase. The consensus phase totally orders the batch attestation shares and create signed block headers, which are subsequently delivered to \emph{assemblers}, who would construct the full block based on the block headers received from the consensus phase, and the actual corresponding transactions retrieved from batchers. The latter phase is called \emph{Block Assembly}, and completes the Arma protocol.

In the next section we define key terms that will be used throughout the rest of this paper, while in the subsequent sections we proceed with the detailed overview of Arma protocol phases and considerations within. 

Afterwards, we deep dive into each of the four stages and explain various considerations in the design.
Finally we discuss operational aspects such as onboarding new parties into an Arma system.

\subsection{Terminology and Definitions}

\noindent We will be using the following terms in the Arma protocol:
\subsubsection{Nodes:} Nodes can be either physical or virtual machines deployed to fulfil a dedicated role, or processes that share their physical machine with other nodes but operate independently of other nodes. 

\subsubsection{Parties:} Parties refer to the Arma consensus participants. Each node in the system is associated to a single party, that deploys and maintains the node. We denote the total number of parties in the system as $N$. 

\subsubsection{Byzantine node/party:} A node or party that is not necessarily abiding by the protocol, but exhibits an arbitrary behavior. Byzantine \cite{ByzGen} nodes can not only crash; a byzantine party may force their nodes to divert from the specified protocol. Byzantine nodes/parties typically represent nodes or parties that have been compromised. In what follows we assume there is at most $F$ out of $N$ corrupted parties and that $F < \frac{N}{3}$. 

\subsubsection{Quorum:} A quorum is defined by the smallest subset of distinct parties
which is guaranteed to intersect with another quorum in at least one correct party.
For instance, if $N=3F+1$, a quorum is exactly $2F+1$ parties, as
$\left(2F+1\right)\cdot2-N=F+1$ which contain at least one correct
party as up to $F$ are malicious. A quorum is always at most $N-F$ but can be less in case $F$ is smaller than $\frac{N}{3}$.

\subsubsection{Shards:} In Arma, a shard represents a logical partition of the transaction
space, dividing it into sets of similar sizes. For instance, one possible
approach is to assign the least significant bit of a transaction to
Shard 1 if it is zero, and to Shard 2 if it is one. Sharding transactions allows parallelization of transaction processing.
\subsubsection{Batch:} A set of transactions of a specific shard bundled together to be sent over the network or persisted on disk.

\subsubsection{Batch Attestation Share:} A message signed by a party attesting that a specific batch of a specific shard
has been persisted on disk by that party.

\begin{figure}[h]
	\begin{framed}
		\centering
		\includegraphics[width=9cm]{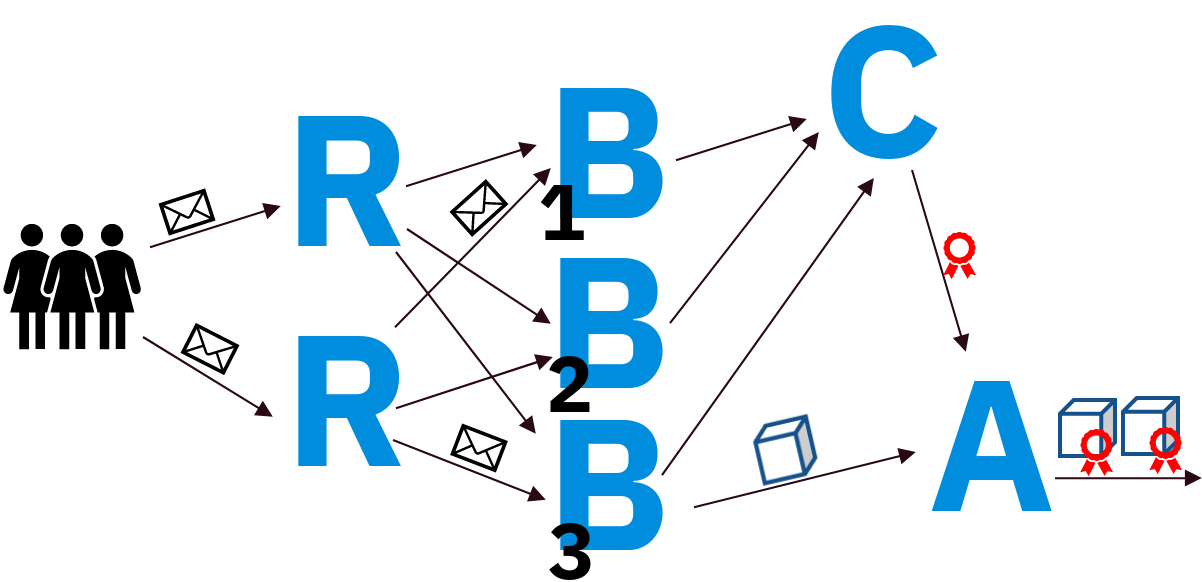}
		\caption{Information flow among components run by each party. Each party runs several (R)outer nodes, (B)atcher nodes (each node designated for a different shard), a single (C)onsenter node and at least one (A)ssembler node. The (B)atcher nodes send batches to the (A)ssembler node and votes on persisted batches to the (C)onsenter node, and the (A)ssembler node receives from the latter signatures over digests of batches voted by the (B)atcher nodes and outputs blocks which are batches with signatures over their digests. 
}%
		\label{fig:party}
	\end{framed}
\vspace{-5mm}
\end{figure}

\subsection{Node roles }
In the Arma system, each party operates one or multiple nodes of different
types: \emph{Router, Batcher, Consensus}, and \emph{Assembler}. In particular, a party's infrastructure would
typically include one or more routers,  multiple batchers (each node designated for a different shard), a single consensus node, 
and at least one assembler. Figure \ref{fig:party} shows the components of a single party in Arma with three shards. 
As mentioned before, the Router nodes check transaction validity and forward the received and validated transactions to the Batchers designated for each transaction's shard. Batchers group received transactions into batches and submit batch attestation shares to Consensus nodes after persisting the transactions locally. Assemblers receive an ordered list of signed batch headers from the Consensus nodes, and retrieve the corresponding batches from 
Batchers. 

As mentioned before, an Arma node can be deployed in its own machine, or can be co-located with
several other nodes in the same machine. More specifically: 

\begin{enumerate}
\item \textbf{Router}: Clients submit transactions to \emph{routers} of multiple parties. 
Routers operate in a stateless manner, and each of them forwards transactions received 
from clients to the batchers of the same party as the router in accordance to their designated shards. 
If required, the router node also verifies that transactions it receives are well formed and 
properly signed by authorized clients. 
\item \textbf{Batcher}: A batcher receives forwarded transactions from the
router(s) and store them in its \emph{memory pool}. At this point, the transaction 
is considered as submitted to the party that operates the batcher by the corresponding client.
A batcher can be either a \emph{primary batcher} or a \emph{secondary batcher}:
\begin{itemize}
\item A primary batcher persists submitted transactions to disk and bundles them into a batch. 
It subsequently forwards the formed batches to the secondary batchers. For each shard there is exactly 
one primary batcher. 
\item A secondary batcher pulls batches from the primary batcher and persists
them to disk. For each shard, there exactly one batcher from each party, except from the party that runs the primary batcher.
\end{itemize}
Upon persistence of a transaction by a primary, the transaction is
removed from its memory pool. Similarly, once a transaction enters a secondary
batcher node, it remains in its in-memory pool until it receives a
batch containing this transaction from the primary. Eventually batchers would submit batch attestation shares to consensus nodes to be totally ordered.
%Each shard contains batcher nodes
%from different parties, and each party may only contribute a single
%batcher node to each shard. 
\item \textbf{Consensus}: The consensus nodes collectively agree on the
total order of transactions by inducing an ordering among the batches
of various shards. They do it by running a Byzantine Fault Tolerant (BFT)
consensus protocol which eventually assigns each batch (from each shard) a 
unique sequence number across all shards. The consensus nodes output a series of batch headers, each signed by a quorum of consensus nodes.
\item \textbf{Assembler}: An Assembler compiles the total order of transactions submitted to 
Arma, by combining the output of consensus nodes, i.e., the total order of batch headers, with 
batches it retrieves from the corresponding batchers to form blocks. The assemblers persist the blocks on disk and can be seen as archivists of the system, as they contain both information of the consensus nodes 
and of the batcher nodes.
\end{enumerate}

%Now when the various nodes within the Arma system are explained, we can map them to the four stages of the Arma consensus protocol as depicted in Figure \ref{fig:stages2} :
%
%\begin{figure}[h]
%	\begin{framed}
%		\centering{{\includegraphics[width=12cm]{stages2} }}%
%		\caption{The node roles that take part in the stages of the Arma consensus protocol. }%
%		\label{fig:stages2}
%	\end{framed}
%\end{figure}

Now that the stages of the Arma consensus protocol are explained, and the various roles of the nodes are defined, we take a deep dive into each of the four stages of consensus.

\subsection{Routing and validation}
In Arma, a client sends her transaction to all parties. Specifically, the transaction
is dispatched to the routers of each party.

A router node is the simplest component among the four, operating in
a stateless manner with two roles: 
\begin{itemize}
	\item Mapping transactions to shards and forwarding them to the corresponding batcher nodes of one's party.
	\item Performing validity checks on the transaction and dropping it if it is found invalid. 
\end{itemize}

As such, a router can be a stateless component and multiple instances of it can be added as needed, 
making transaction validation and routing horizontally scalable.

\subsubsection{Mapping transactions to shards}

To ensure censorship resistance and de-duplication, it is essential that transactions are mapped to shards in a deterministic manner.
In this way, assuming a client follows the protocol, and submits the transaction to all parties, all honest parties' routers would  forward the received transactions to the appropriate batchers, ensuring that all honest batchers of the shard will receive the transaction.
Additionally, for effective load balancing across the shards, the router must distribute the transactions
among the known $k$ shards in a manner that is as close to a uniform distribution
across the set $\left\{ 1,..,k\right\} $ as possible.

While a cryptographic hash function could serve as a suitable choice,
it is worth noting that the router does not require the collision
resistance and one-way properties. Instead, more efficient alternatives
like a CRC checksum can be employed without compromising its functionality.

\subsubsection{Transaction verification}

Besides forwarding transactions to their corresponding shards, routers also ensure only transactions that are well formed and properly signed by clients are forwarded. In earlier protocols such as \cite{Mir-BFT}, each party ran a single node which was involved in verifying all transactions arriving from clients to that party. In Arma, however, a party can utilize the routers to horizontally scale the CPU intensive task of verifying signatures and alleviating the batcher nodes. Identifying ineligible transactions and excluding them from the next steps, ensure that system resources are properly devoted to the processing of correctly formed transactions.

\subsection{Transaction Batching}
\label{sec:batching}
In the transaction batching phase transactions are delivered to batchers and bundled together into batches, in a way that ensures that (i) transaction batches persist safely into stable storage for redundancy and later retrieval, and (ii) headers of transaction batches are submitted to the consensus phase. 
Recall that batchers are grouped into shards. Each party runs a single batcher for every shard in the system, that can act either as a primary batcher, or as a secondary one. Each shard has a single designated primary batcher node, and the rest are secondaries.

\begin{figure}
	\begin{framed}
		\centering
		\subfloat[\centering A transaction sent from a client is sent to all router nodes and is then dispatched to the batcher node of each party]{{\includegraphics[width=5cm]{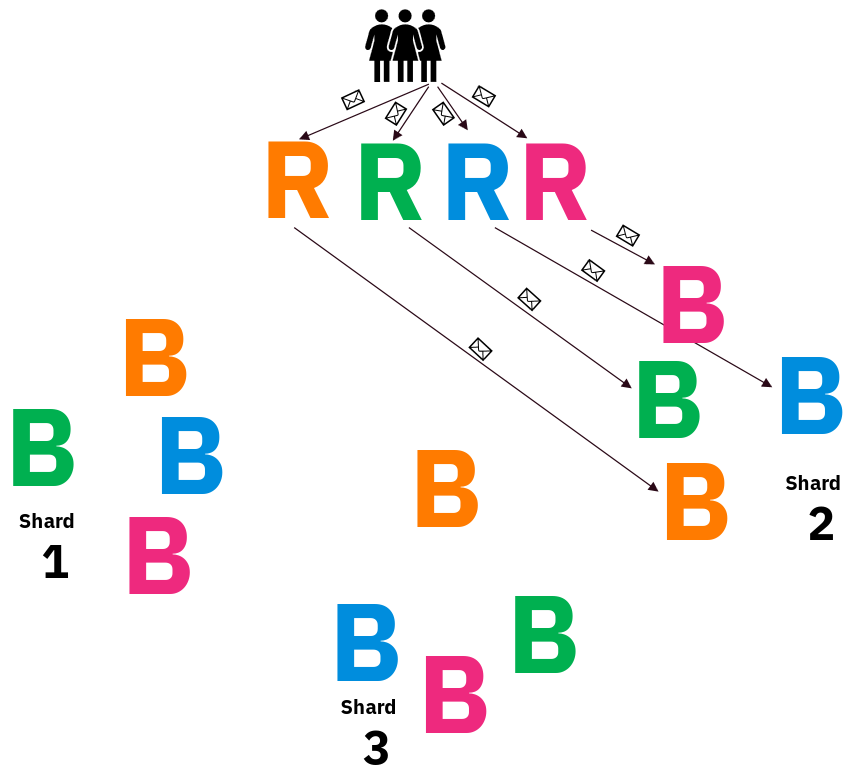} }}%
		\qquad
		\qquad
		\qquad
		\subfloat[\centering In each shard, the primary batcher node broadcasts a batch of transactions to the secondary batcher nodes]{{\includegraphics[width=5cm]{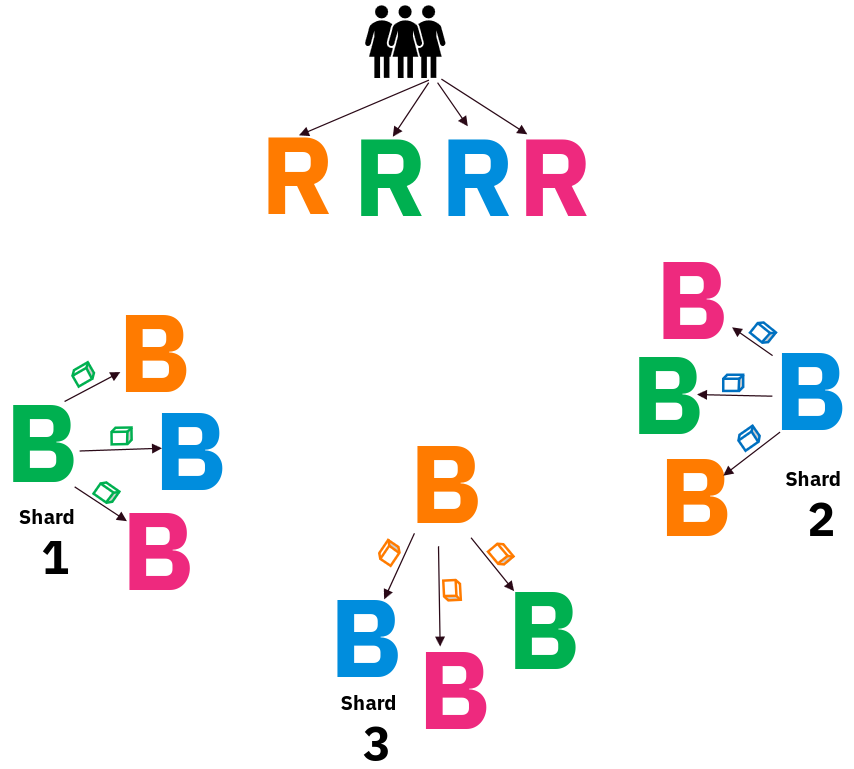} }}%
		\caption{A birds eye view of router and batcher nodes of all parties (a) routing a transaction to the appropriate shard, and (b)  broadcasting transaction batches}%
		\label{fig:routingbatching}
	\end{framed}
\end{figure}

\begin{figure}[h]
	\fbox{
		\begin{minipage}[t]{0.5\linewidth}
			\vspace{0pt}  
			%\fbox{\parbox{0.95\linewidth}{
					\begin{algorithm}[H] \label{alg:primbatcher}
						\KwIn{\\$~~~~~$Private key for signing $sk$,\\$~~~~~$ledger $\mathcal{L}$,\\$~~~~~$Total Order Broadcast $TO$,\\$~~~~~$Transaction memory pool $\mathcal{M}$}
						\While{I am primary}{
							b $\leftarrow \mathcal{M}.Get()$\;
							$\sigma$ $\leftarrow$ sign($sk$, b.Seq || b.Digest || b.Shard || $i$) \;
							$\mathcal{L} \leftarrow \mathcal{L} ~ || ~ b$\;
							$TO.Broadcast\left( \big< \sigma, b.Seq, b.Digest, b.Shard, i \big> \right)$\;	
						}
						\vspace{36mm}
						\caption{Primary batcher $i$}
					\end{algorithm}
					%}}	
		\end{minipage}%
		\hfill\vline\hfill
		\begin{minipage}[t]{0.5\linewidth}
			\vspace{0pt}
			%\fbox{\parbox{0.98\linewidth}{
					\begin{algorithm}[H] \label{alg:secbatcher}
						\KwIn{\\$~~~~~$Private key for signing $sk$,\\$~~~~~$ledger $\mathcal{L}$,\\$~~~~~$Total Order Broadcast $TO$,\\$~~~~~$Stream of batches from the primary $\mathcal{B}$}
						\KwIn{Total Order Broadcast $TO$,\\$~~~~~$Transaction memory pool $\mathcal{M}$}			
						\While{I am not primary}{
							$seq$ $\leftarrow \mathcal{L}.Height()$\;
							b $\leftarrow \mathcal{B}.RetrieveBatch(seq)$\;
							\If{$invalidTxInBatch\left(b\right)$}{
								$\sigma$ $\leftarrow$ sign($sk$, t || b.Shard || $i$)\;		
								$TO.Broadcast\left( \big< \sigma, t, b.Shard, i \big> \right)$\;
								\Return{}	
							}
							$\mathcal{L} \leftarrow \mathcal{L} ~ || ~ b$\;				
							$\mathcal{M}.Remove(b.Requests)$\;
							$\sigma$ $\leftarrow$ sign($sk$, b.Seq || b.Digest || b.Shard || $i$)\;						
							$TO.Broadcast\left( \big< \sigma, b.Seq, b.Digest, b.Shard, i \big> \right)$\;
						}
						\caption{$~~$Secondary batcher $i$ for term $t$}
					\end{algorithm}
					%}}	
		\end{minipage}
	}
	\caption{Pseudocode for batchers (primary and secondary). Once a batch is appended to the ledger (line 4), it will eventually be outputted by $\mathcal{B}$.}
	\label{fig:batchers}
	
\end{figure}

\begin{figure}
	\begin{framed}
		\centering{{\includegraphics[width=10cm]{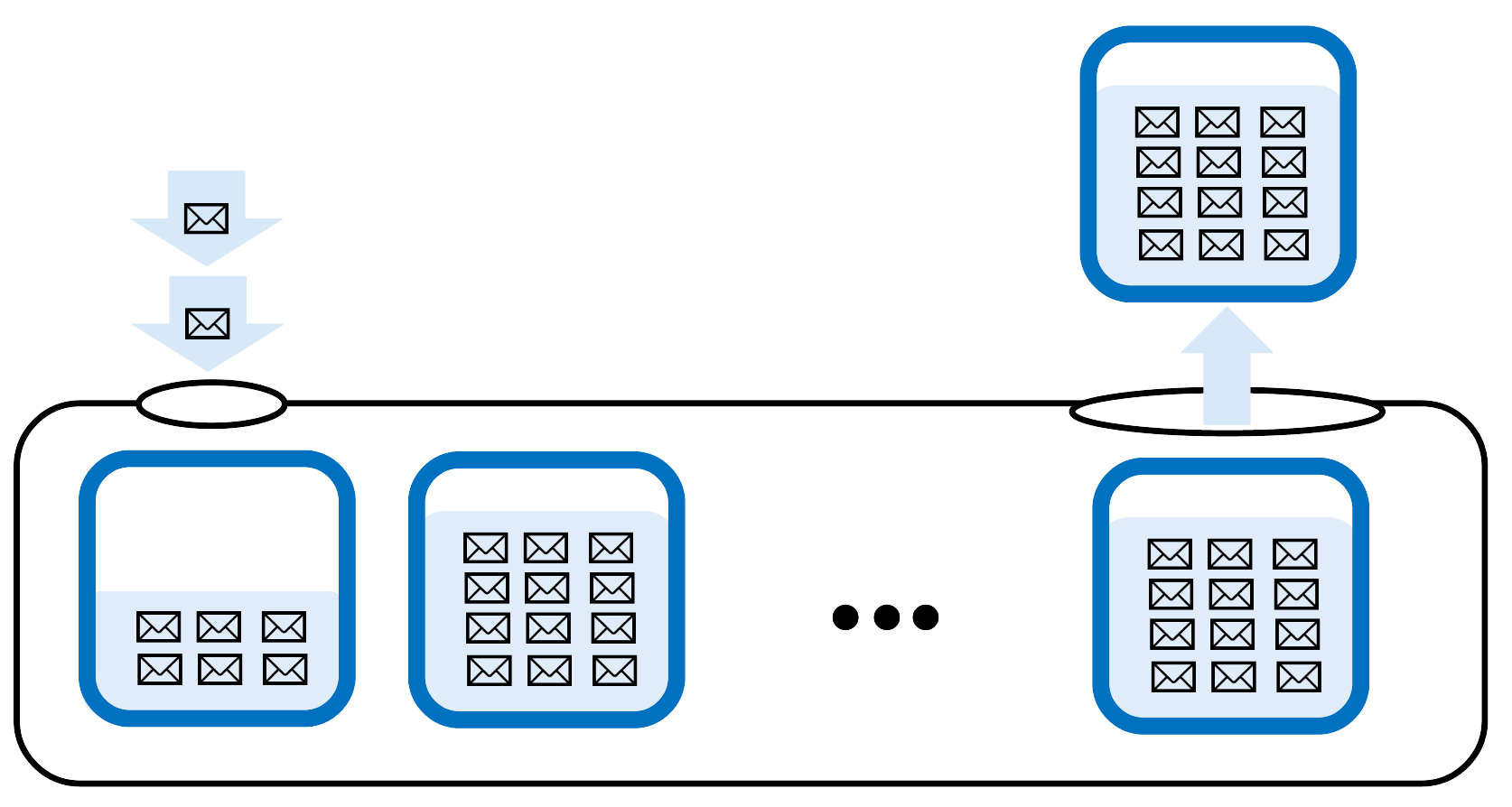} }}%
		\caption{Transaction memory pool for primary batchers}%
		\label{fig:bundling}
	\end{framed}
\end{figure}

\subsubsection{Transaction dissemination}
As mentioned before, a transaction is submitted by a client to all the parties in the system (i.e., their router node). 
Routers subsequently deliver the transaction to their party's batcher for the corresponding shard. 
It is only then, when the client deems this transaction submission as successful to that party.
We assume that a correct client considers a transaction submission as successful when it has been successfully submitted to $N-F$ parties. We further consider each unreachable party as a faulty one. An more relaxed failure model where a party can be unreachable to a correct client but still considered not faulty can be supported by having the client send to $F+1$ parties and have the parties gossip the transactions among each other.

As a transaction arrives from a router node to the primary batcher of the shard the transaction corresponds to, the primary batcher includes it in a batch
and persists the batch to disk. When the transaction arrives from a router node to a secondary node of the same shard, the batcher persists puts 
it into its in-memory pool. The secondary batcher then pulls batches from the primary and after persisting them to disk, removes transactions that appear in the received batches from its in-memory pool (if they have already been added there). Conversly, transactions that have been received from clients shortly after appearing in batches pulled from the primary do not get added to the memory pool, as they have already been processed.

Figure \ref{fig:routingbatching} shows the transaction flow from clients to batchers through the router nodes, and also how batches of transactions are broadcast among batchers in each shard.

Once a batch is persisted to disk, each batcher be it primary or secondary, creates a \emph{Batch Attestation Share} by signing over a message $<shard,primary,sequence,digest>$
where $shard$ and $primary$ are the numerical identifiers of the shard and party of the primary batcher node of the shard respectively, $sequence$ is how many batches were previously observed by the primary batcher node, and $digest$ is a hash or a Merkle tree root of the transactions in the batch, and totally orders the batch attestation share by sending it to all consensus nodes, acting as a client to the consensus nodes. 

It is the responsibility of each batcher of a shard to ensure that the batch attestation share has been sent to all possible consensus nodes, which is at least a quorum.
Figure \ref{fig:batchers} shows the pseudocode for batcher nodes in both primary and secondary roles, where sending the batch attestation share to consensus nodes is abstracted by a Total Order primitive.

\subsubsection{Criteria for inclusion of a batch into total order}
\label{sec:inclusion-batch-attestations}
Each shard has a total of $N$ parties, and at least a quorum of batch attestation
shares is collected by having them totally ordered. Once a $F+1$
distinct shares for the same batch attestation share, i.e., triplet $<primary,sequence,digest>$, 
are totally ordered, it means that the associated batch can be retrieved by 
at least one correct party.

%\sout{However, if a quorum of shares were collected,
%but no $F+1$ shares for the same $<primary,sequence,digest>$
%are found among them, it proves that the primary batcher has equivocated
%its batch.}\footnote{\sout{Assume in contradiction the primary hasn't equivocated. Then up to
%	$F$ parties claim they received batch $b$ and up to $F$ claim they
%	received $b'\neq b$ and at least one 1 party claims to receive batch
%	$b''\notin\left\{ b,b'\right\} $, therefore at least $F+1$ parties
%	lie about the batch they receive, which is a contradiction to the
%	assumption of up to $F$ parties being discorrect.}} 

\subsubsection{Efficient transaction bundling}

Given that a batcher node can function as either a primary or secondary,
it is necessary to prioritize efficiency in transaction bundling.

The challenge lies in swiftly retrieving transactions from the memory
pool while maintaining the order of their arrival and simultaneously
allowing uninterrupted insertion of new transactions into the pool.

The Arma memory pool utilizes a mechanism where the retrieval of batches
from the memory pool has a time complexity of O(1). This is achieved
by having transaction insertions and batch retrievals not be conflicting with each other.
At any given time, there is a pending batch being filled and a queue
of full batches awaiting dispatch. When a batcher node retrieves a
batch from the memory pool, it simply dequeues the oldest full batch
from the queue. If no full batch is present, it retrieves the pending
batch that is currently being filled. As transactions enter the memory
pool, they contribute to filling the latest batch, and once the batch reaches
a certain size, it is atomically enqueued into the queue of full batches
while an empty batch takes its place. Figure \ref{fig:bundling} depicts how transactions enter the memory pool (left side) and fill batches until they are full to form a queue. The next batch to be proposed is the oldest batch created which is on the right side.

\subsubsection{Tracking transactions by secondary batcher nodes}

The primary batcher's role is to quickly bundle a set of transactions,
enabling the secondary batchers to retrieve them efficiently. On the
other hand, the secondary batchers are responsible for tracking the
transactions in their memory and promptly detecting if any transactions
are not being dispatched within the expected timeframe. 

Before we delve into the various mechanisms in which Arma ensures transactions
that are correctly sent by a client are eventually totally ordered,
we outline how Arma secondary batchers detect that transactions have not been sent by primary batcher nodes.

\begin{figure}[h]
	\begin{framed}
		\centering{{\includegraphics[width=10cm]{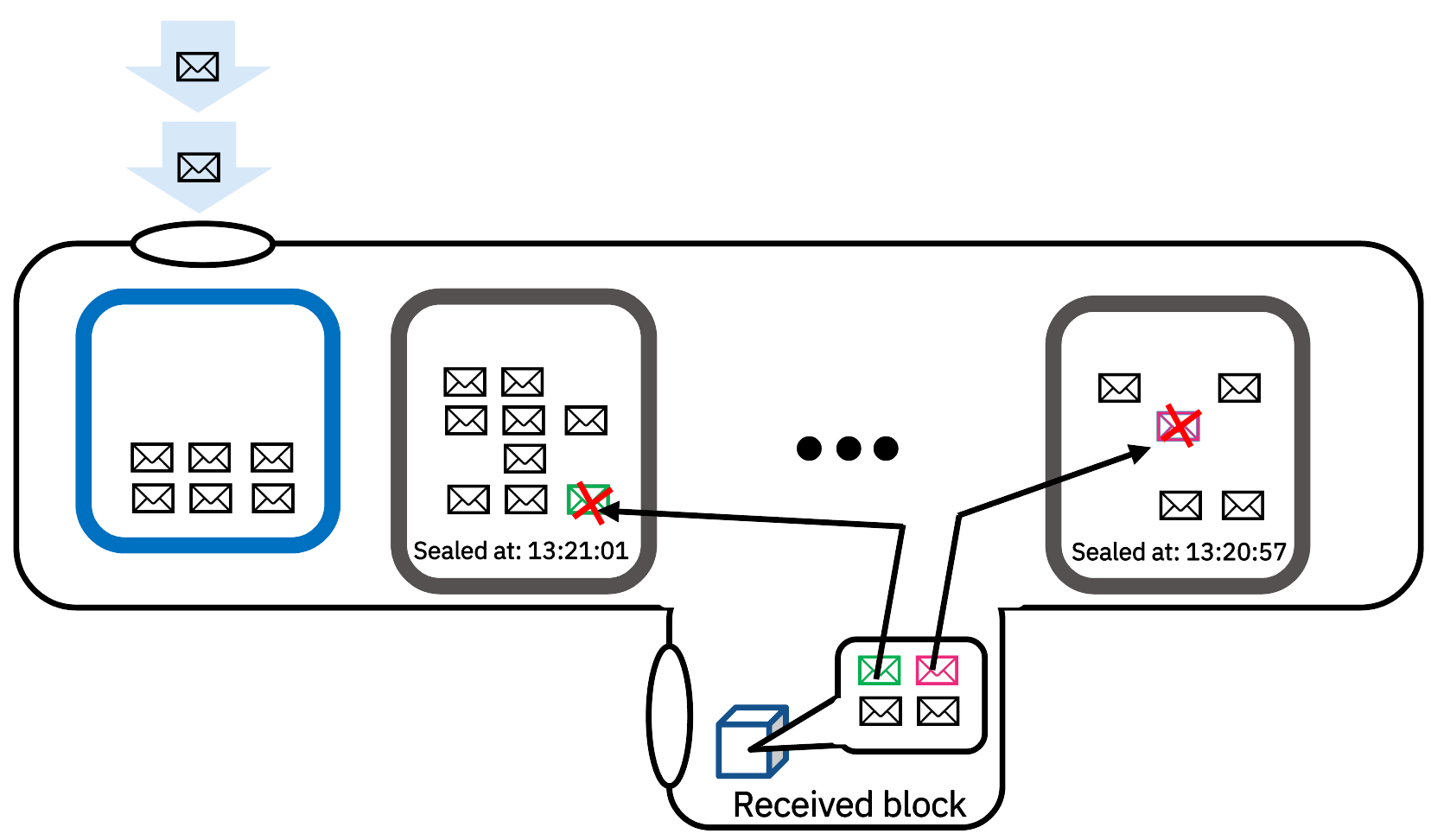} }}%
		\caption{Transaction memory pool for secondary batchers}%
		\label{fig:tracking}
	\end{framed}
\end{figure}

Efficient detection of transaction censorship involves recording the
entry time of transactions into the secondary batcher node's memory
pool and identifying instances where transactions remain in the pool
for an extended period. The key concept in achieving efficient censorship
detection is that precision is not crucial in the case of actual censorship.
In the Arma memory pool, incoming transactions are inserted into specific
buckets. These buckets are assigned timestamps periodically and are
subsequently sealed, preventing further transaction insertions. When
transactions arrive from the primary batcher node, they are removed
from the corresponding buckets where they were initially inserted.
A sealed bucket that remains non-empty for an excessively long duration
indicates either the transactions within it did not reach the primary
node or that the primary node is censoring the transactions. Garbage
collection occurs for sealed and emptied buckets, while sealed yet
non-empty buckets serve as indicators of potential censorship or transmission
issues. Figure \ref{fig:tracking} depicts how transactions are tracked in a secondary batcher's memory pool.

\subsubsection{Eventual total ordering of transactions}

Arma provides the guarantee that if a client submits its transaction
to all parties, it will eventually be included in a block without
the need for the client to retry the submission. It is important to
note that in the Byzantine setting in which Arma operates, nodes controlled
by malicious parties have the potential to deviate from the protocol.

For instance, an Arma primary batcher node may encounter a failure
or intentionally disregard client transactions. Similarly, a consensus
leader node has the ability to ignore batch attestation shares
sent by batcher nodes or experience a crash. These scenarios highlight
the challenges and risks associated with Byzantine behavior, which
Arma aims to address through its consensus protocol and the mechanisms
it employs. Despite these challenges, Arma ensures that, over time,
all valid client transactions will be included in a block without
requiring the client to reattempt the submission.

\subsubsection{Censorship resistance in Byzantine Fault Tolerant protocols}

A Byzantine Fault Tolerant (BFT) consensus protocol that ensures censorship
resistance ensures that each transaction is eventually included in some finalized block. 
A common approach involves non leader (follower) nodes
recording the time when a transaction is received. If a transaction
is not included in a block sent by the leader within a specific time
period, the follower nodes forward the transaction to the leader.
If the leader still does not include the transaction in subsequent
blocks, a view change protocol is initiated. During a view change,
the leader role is rotated, and one of the follower nodes becomes
the new leader. 
Another technique of ensuring censorship resistance is periodical
leader rotation. By periodically switching the leader role, a transaction
censored by a malicious leader node is eventually included in a block
once an correct node becomes the leader.

\subsubsection{Censorship resistance in Arma batchers}
\label{sec:censorship-resistance}
In Arma, a similar technique to the aforementioned censorship resistance
method is employed to ensure that batcher primary nodes do not censor
transactions from clients. However, instead of initiating a view change
protocol among the batcher nodes, Arma utilizes a complaint voting
mechanism.

In Arma, if a batcher secondary node receives a transaction from a client and it has not been included in a block sent by the primary batcher node for a period of time, it sends the transaction to the primary node itself. This is to ensure that the primary batcher received the transaction. Otherwise, a faulty client may have not sent the transaction to the primary batcher node and it will be falsely accused of censorship.
Only when the transaction is still not received in a batch from the primary batcher node after sending it directly, the secondary node suspects the primary node of censorship.
The secondary batcher node then sends a complaint vote to all consensus nodes, expressing
its concern against the suspected primary batcher. These complaint
votes are totally ordered alongside the batch attestation shares.
Once a threshold of $F+1$ complaint votes is gathered against a specific batcher
primary, the batcher nodes collectively designate the next secondary
batcher node as the new primary batcher for the respective shard.

\subsubsection{Verifying batches sent from primary batcher nodes}
As previously mentioned, part of the censorship resistance mechanism of batchers is having the secondary batcher nodes
send transactions that were not included in a batch sent from the primary batcher.
This is done by the secondary batcher sending the transaction to the router node of the primary batcher's party. 
Hence, it is ensured that that secondary batcher nodes cannot send bogus transactions and compete with authentic ones over space in the in-memory transaction pool.

It is left to prevent malicious primary batcher nodes from sending bogus transactions in their batches sent to the secondary nodes. While such transactions do not end up in the in-memory transaction pool of the secondary nodes, as they came from the primary node, it still should be avoided as it needlessly wastes resources and lowers the effective throughput of the system, as they are sent over the network and written to disk. The overwhelmingly common approach employed by existing systems is to have the nodes verify the transactions sent from the primary batcher node. While this approach indeed prevents inclusion of bogus transactions, it is very resource intensive and does not scale horizontally as the secondary node is a single machine \footnote{Of course, the node may delegate the verification to several machines and then aggregate the results, but this requires to send the batch and wait for results, which carries a latency overhead} like transactions received by router nodes. To that end, Arma employs a probabilistic approach: Every secondary batcher node chooses a random subset of transactions to verify in each batch received from the primary batcher node. 
More specifically, each correct secondary batcher node picks $R$ transactions to verify among the $M$ transaction in a batch with $K$ invalid transaction.

%In order to miss all invalid transactions, the correct batcher node needs to pick any of the ${M-K \choose R}$ combinations among the ${M \choose R}$ combinations. Since there are a total of $N-F$ correct secondary batchers and their trials are independent, each batch has a probability of $\left( \frac{{M-K \choose R}}{{M \choose R}} \right)^{N-F}$ to be undetected. \textcolor{orange}{Yacov: I will move this to security analysis once I have it.}
%Even for a single batch and a single secondary batcher node, the probability of a batch with 50\% invalid transactions being undetected is slim. To understand why, imagine a block of 5000 transactions and a secondary batcher picking randomly 50 of them to verify, which is only one percent. Since 50 is much smaller than 5000,\kaoutar{The fact that 50 << 5000 is irrelevant to what follows.} the secondary batcher picking only valid transactions is similar to 50 coin toss experiments that need all to have the same outcome.

If one of these randomly chosen transactions are found to be invalid, the secondary batcher node's party then issues a complaint to overthrow the primary node and make a different party be the primary for that shard. 

%While a malicious primary batcher node can send some batches containing bogus transactions, the probability of it being detected grows exponentially\kaoutar{with what? it should be mentioned that probability increases exp. with the number of checked transactions.} and converges to 1 as it creates new batches\kaoutar{Why would this matter. A malicious batcher will attempt to make the events independent}. Hence, such a denial of service attack by a malicious primary can only be done for a limited period of time at the cost of verifying only a small subset of the transactions in each batch. \kaoutar{there should be a quantification of the number of transactions to be checked randomly by each batcher. One starts with the tolerated fraction of wasted resources, let's say $10\%$, fixing the probability of evading detection, and then computing the number of transactions to be checked. That number is then divided by $2F+1$ and that give how many transactions one needs to check per batch.}

\subsubsection{Determining the complaint threshold to rotate the primary batcher node}
Effectively, the batcher nodes utilize the BFT consensus protocol
executed by the consensus nodes as a bulletin board that tracks, for
each shard, which party runs the primary batcher node for a given term. This approach
facilitates agreement on the misbehavior of faulty primary batcher nodes, and implictly provides auditability on the misbehavior.

The threshold of complaint votes that induces a change in the primary batcher node is $F+1$ complaints from distinct nodes about a specific term. When $F+1$ complaint votes are collected, the term is incremented by 1 and the next batcher defined is then made the primary for the shard. By defining the threshold to be $F+1$, we are assured that a correct party considers the primary batcher node of a shard to be faulty, and that $F$ faulty parties are not enough to falsely accuse a correct primary batcher node. Moreover, this ensures that if $F+1$ correct nodes cannot pull batches from the primary batcher node, then the primary batcher node is replaced in a timely manner.

\subsubsection{Primary batcher failover}

In a Byzantine Fault Tolerant (BFT) consensus protocol, both view
changes (changing the leader node) and regular blocks that are finalized through consensus are
totally ordered with respect to each other. This means that if a block
$b$ is broadcasted by the leader of view $v$ and subsequently finalized,
and later a view change protocol occurs, incrementing the view to
$v+1$, all correct nodes possess knowledge of whether block $b$
was finalized as part of view $v$ or as part of the view change protocol.

However, in the Arma protocol, the retrieval of batches by secondary
batcher nodes from the primary batcher is independent of the retrieval
of updates from the consensus. As a result, the progress of batches
operates asynchronously with respect to the rotation of the primary
batcher for a shard. This introduces a potential problem where a transaction
might be included in a batch of a primary batcher, persisted across
several batchers, and subsequently an update is received from the
consensuss designating a different primary batcher. If the corresponding
batch attestation shares are discarded, the transactions associated
with that batch could be lost. It is not guaranteed that sufficient
batch attestation shares can be collected for that batch.

To address these scenarios, concrete rules which are discussed next, are required to determine
how the rotation of primary batchers should be interwoven with deciding which batches proposed by previous primary batchers are to be proposed again by the newly appointed primary batchers. Establishing
strict rules for primary batcher rotation helps to ensure the integrity
and consistency of batches and their corresponding attestation shares,
minimizing the risk of transaction loss or inconsistency during the
rotation process. 

When rotating a primary, the protocol should ensure that transactions
are not dropped due to the rotation. The challenge is that the transaction
memory pool in a batcher deletes a transaction once a batch containing
it was received from a primary. Thus, a secondary batcher might become
a primary after deleting a transaction that came in a batch from a
primary that is no longer a primary. More formally, consider a batch
$b^{p}$ sent from primary $p$. If $p$ crashes, a new primary needs
to take its place. When a secondary batcher $q$ is to become the
primary instead of $p$, it may or may not have received $b^{p}$.
We split into cases according to two events: (A) Whether $b^{p}$
was received by $q$; and (B) whether $b^{p}$ was received by a quorum
of parties.
\begin{enumerate}
\item $AB$: $b^{p}$ was received by a quorum of parties and by $q$. Since
$q$ received $b^{p}$ it will delete its transactions from its memory pool and
then $F+1$ batch attestations will be eventually ordered
by consensus, since a quorum of parties received $b_{p}$.
\item $\bar{A}\bar{B}$: $b^{p}$ was not received by a quorum of parties
and neither was received by $q$. Then, every transaction in $b^{p}$
is still in the memory pool of $q$ and will be included only once
in a batch made by $q$. If $q$ is malicious and ignores the transactions in $b^{p}$ we split into
cases according to how many batchers received $b^{p}$: 
\begin{enumerate}
	\item If at least $F+1$ (but less than a quorum) correct batchers received $b^{p}$ it means $F+1$
	batch attestation shares for $b^{p}$ will eventually be totally
	ordered by the consensuss. 
	\item Else, $F$ or less correct batchers received $b^{p}$. Since an honest client submits its
	transaction to all parties out of which up to $F$ are faulty, the transactions of $b^{p}$ reached $N-F$ correct parties.
	Therefore, at least $F+1$ correct batchers have the transactions of $b^{p}$ (since $F$ or less correct batchers received $b^{p}$).
	Thus, it follows from the censorship resistance mechanism that if $q$ will not propose batches containing the transactions
	of $b^{p}$, enough complaints will be gathered to make $q$ no longer
	be the primary batcher, in which case eventually a correct batcher will be chosen.
\end{enumerate}
\item $\bar{A}B$: $b^{p}$ was received by a quorum of parties but not
by $q$. Then, each transaction in $b^{p}$ is still in the memory
pool of $q$ and will be included in a batch made by $q$. The transactions
will therefore may be included twice - once in a batch made by $p$ and
once in a batch made by $q$. 
\item $A\bar{B}$: $b^{p}$ was not received by a quorum of parties, but
was received by $q$. Since $q$ is the new primary and it has $b^{p}$,
it then proposes its own batch $b^{q}$ containing only the transactions
of $b^{p}$ as depicted in Algorithm \ref{fig:failover}. As the batch attestation shares of $b^{q}$ are totally
ordered by consensus, either $F+1$ batch attestation shares
of $b^{p}$ or of $b^{q}$ will be collected. However, $q$ might
be malicious and decide to ignore $b^{p}$, which is covered in case (2) $\bar{A}\bar{B}$.
\end{enumerate}

\begin{figure}[h]
	\fbox{
		\vspace{0pt}
		\begin{algorithm}[H]\label{fig:failover}
			\KwIn{\\$~~~~~$An ordered set of pending batch attestation shares
				from earlier iterations $P=\{bas_1, bas_2, ..., bas_m\}$\\
				$~~~~~$Total Order Broadcast $TO$\\$~~~~~$Ledger $\mathcal{L}$}
			$T \leftarrow \emptyset$ \; 
			$C \leftarrow \emptyset$ \; 
			{
				\ForEach{$bas\in P$}{
					$k \leftarrow < bas.Seq, bas.Shard, bas.Digest \big>$ \; 
					$C[k] \leftarrow C[k]+1$ \;
					$T[k] \leftarrow T[k] \cup bas$ \;
				}
				
				\ForEach{$k\in C$}{
					\If{$C[k] < F+1$}{
						$T \leftarrow T \setminus T[k]$ \;		
					}
				}
				
				$P \leftarrow P \setminus T$ \;	
				$S \leftarrow \emptyset$ \;					
				
				$seq \leftarrow 0$ \; 
				
				\ForEach{$bas\in P$}{
					$b \leftarrow \mathcal{L}.Retrieve\left(bas\right)$\;
					\If{$b \neq \bot ~ \wedge ~ S[b.Digest] \neq \bot$}{
						$b.Primary \leftarrow i$\;						
						$b.Seq \leftarrow seq$\;
						$\mathcal{L} \leftarrow \mathcal{L} ~||~ b$\;						
						$seq \leftarrow seq+1$\;	
						$S[b.Digest] \leftarrow 1$
					}
				}					
			}
			\caption{$~~$Pseudocode for batcher failover, run by the new primary batcher $i$. Lines 1-13 determine which batches proposed by previous primary batchers need to be carried over by this batcher. For simplicity, in line 14 we assume this was the first time this batcher is primary for this shard. Lines 15-23 propose these batches again.}
		\end{algorithm}
	}	
\end{figure}

\subsubsection{Decentralized batch attestation share broadcast}
As in the work of \cite{Narwhale}, Arma totally orders messages attesting that a specific batch of transactions has been safely persisted on some predefined subset of parties. However, Arma does this in a decentralized and trust-less manner.
While in \cite{Narwhale}, the primary batcher node collects the attestations from the rest of the nodes, in Arma, each 
batcher node is responsible of sending its own batch attestation share to the consensus nodes. 
While at a first glance this design choice may seem as an inefficient one, as it linearly amplifies the number of messages totally ordered by consensus, it actually gives the system more robustness.

If the primary batcher node crahes, or is misbehaving, it may not be totally ordering batch attestation shares at all.
In such a case, a heuristic mechanism which monitors the output of the consensus nodes and reports to the batcher nodes is required.
Unfortunately, constructing such a mechanism is difficult because it is not clear why an attestation hasn't been totally ordered:
It can either be because the primary batcher node is malicious, or it can be because the consensus leader is being malicious. 
In order to avoid false positives in the reporting mechanism, the timeout for the reporting mechanism must be higher than the censorship resistance mechanism for the consensus by a large enough margin, otherwise the consensus leader can be wrongly suspected. 
Moreover, relying on the correctness of the primary batcher to submit attestations created by the secondary batcher to consensus to be totally ordered, opens the protocol for a denial of service attack: A primary batcher node controlled by the adversary can create new batches but not totally order the corresponding attestations until it is suspected by the secondary batcher nodes.
The secondary batcher nodes now need to reconcile amongst themselves the batch attestation shares they have created in the meantime. After reconciliation ends, they ought to send the batch attestation shares by at least $F+1$, lest they rely on malicious batcher nodes once more. However, this means that a sudden load is now imposed on the consensus protocol.
This is in contrast to the way Arma works, where batch attestation shares are totally ordered immediately and by all batcher nodes at a steady rate.

Primary batcher nodes from all shards produce batches of transactions which are then received by secondary batcher nodes. The batches are produced and disseminated in parallel, and as such, parties cannot tell whether a batch of shard $i$ precedes or succeeds a batch of shard $j$. We next show how this problem is solved by totally ordering the corresponding batch attestation shares by the consensus nodes.

\subsection{Inducing total order across batches via BFT consensus}
As explained previously, once a batcher node persists a batch, it sends a batch attestation share to all consensus nodes to be totally ordered by having consensus nodes participate in a Byzantine Fault Tolerant (BFT) consensus protocol. For simplicity, we say that the BFT protocol advances in rounds, and in each round it totally orders some batch attestation shares. The rounds do not refer to communication rounds or stages in the consensus protocol, but rather to proposals sent from the leader that have been finished being ordered by the BFT consensus protocol.

\subsubsection{Collecting batch attestation shares}
\label{sec:collecting-batch-attestation-shares}
\begin{figure}[h]%
	 \begin{framed}
	\centering{{\includegraphics[width=15cm]{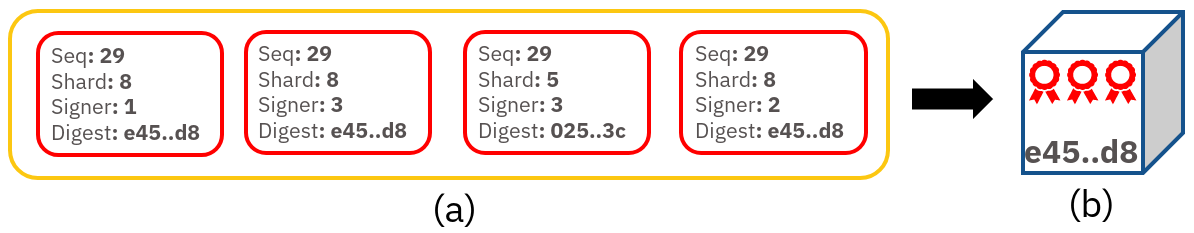} }}%
	\caption{(a) Batch attestation shares totally ordered in a consensus round, for which (b) a corresponding block header with a quorum of signatures is assembled}%
	\label{fig:bas}
	 \end{framed}
\end{figure}

Each batcher, whether primary or secondary, sends a batch attestation
share upon successfully persisting a batch on disk. The batch attestation shares 
are totally ordered by consensus in some consensus round, and during the
consensus protocol, the Arma consensus nodes execute a non-interactive computation
which determiness for which transaction batches enough batch attestation shares have been
collected. Once $F+1$
number of batch attestation shares for a specific transaction batch are totally
ordered, the consensus nodes collaborate to assemble a quorum of signatures
over a block header corresponding to that batch, as depicted in Figure \ref{fig:bas} .
the signed block header which contains a hash
pointer to the previous block header, extending the block header chain.
Each block header contains a digest of its corresponding batch, which
is either a hash or a Merkle root of the batch. While it is possible
to sign each batch, a more efficient method is to create a block header
with multiple digests, each corresponding to a batch. This amortizes
signature verification over several batches, making it cheaper to
verify a block.

\begin{figure}[h]
	\fbox{
		\vspace{0pt}
		\begin{algorithm}[H]\label{fig:consensuss}
			\KwIn{\\$~~~~~$An ordered set of pending batch attestation shares
				from earlier iterations $P=\{bas_1, bas_2, ..., bas_m\}$,\\
				$~~~~~$A batch containing  batch attestation shares sent from the leader consensus node $B=\{bas_1, bas_2, ..., bas_k\}$}
			\KwOut{\\
				$~~~~~$An ordered set of pending batch attestation shares for future iterations $P$,\\
				$~~~~~$An ordered set of sets of distinct $F+1$ batch attestation shares $T$.
			}
			$P \leftarrow P \cup B$ \;
			$T \leftarrow \emptyset$ \; 
			$C \leftarrow \emptyset$ \; 
			{
				\ForEach{$bas\in P$}{
					$k \leftarrow < bas.Seq, bas.Shard, bas.Digest \big>$ \; 
					$C[k] \leftarrow C[k]+1$ \;
					$T[k] \leftarrow T[k] \cup bas$ \;
				}
				
				\ForEach{$k\in C$}{
					\If{$C[k] < F+1$}{
						$T \leftarrow T \setminus T[k]$ \;		
					}
				}
				
				$P \leftarrow P \setminus T$ \;	
				
				\KwRet{P, T}
			}	
			\caption{$~~$Pseudocode for processing batch attestation shares}
		\end{algorithm}
	}	
\end{figure}

\subsubsection{Collecting thresholds of batch attestation shares}
During the process of totally ordering batch attestation shares,
a very likely scenario is one where $F+1$
batch attestation shares is totally ordered, but in different BFT rounds. To handle
this situation, Arma maintains a list of pending batch attestation
shares within the metadata section of each BFT round. Once $F+1$ batch attestation shares corresponding to a batch are collected, the consensus nodes know they can collaborate to create a block header corresponding to the batch, include it in the hash chain and sign over it.
In Algorithm \ref{fig:consensuss} it is explained how thresholds of $F+1$ batch attestation shares for the same batch are grouped together and extracted even though they arrive in different consensus agreement rounds. 
The batch attestation shares agreed upon in a round of BFT are input into Algorithm \ref{fig:consensuss} and the output from it is then passed as input in the next BFT round. 

The reason for this approach is to ensure all consensus nodes would process batch attestation shares deterministically and thus sign over the same block headers comprising the hash chain. However, there is a non trivial challenge in this design: After a $F+1$ distinct batch attestation shares are accumulated in the pending list, they are purged by Algorithm \ref{fig:consensuss} , but additional batch attestation shares may then accumulate there in future rounds. This is because more than $F+1$ batcher nodes submit their batch attestation fragments, but there is no guarantee they end up agreed upon in the same BFT round. For example, in a four node setting with a single faulty node, the threshold $F+1$ is two, however three batch attestation shares are created. We denote a batch attestation share that is totally ordered by consensus in a later round than the round which collected a $F+1$ batch attestation shares as an \emph{orphaned} batch attestation share. Clearly, an orphaned batch attestation share will never be purged from the pending list by Algorithm \ref{fig:consensuss}. Therefore as the system processes new transactions, the pending list will contain more and more orphaned batch attestation shares.

Another problem is that too many batch attestations accumulated may lead to creation of two block headers with the same batch digest: For example, in a system of four nodes, two thresholds of $F+1$ attestation shares may be collected by two pairs of two batch attestation shares.

\begin{figure}[h]
	\begin{framed}
		\centering{{\includegraphics[width=17cm]{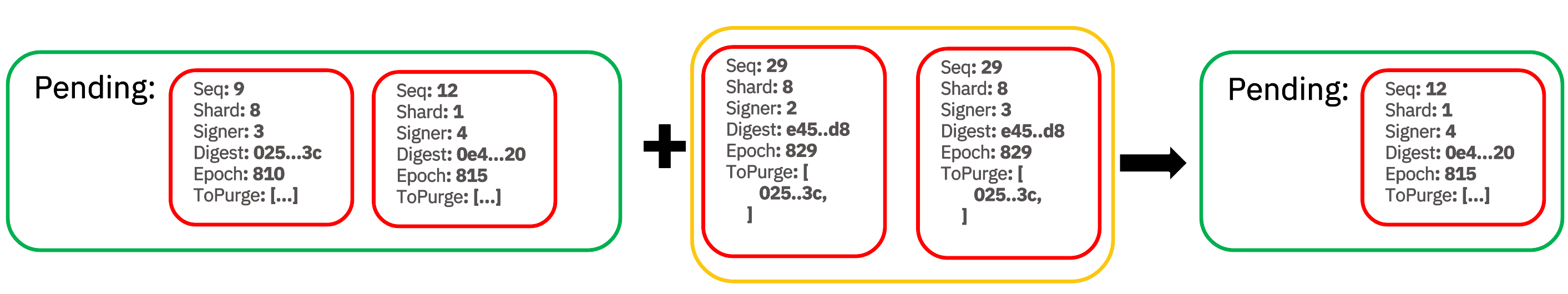} }}%
		\caption{Purging orphaned batch attestation share of transaction batch with digest 025...3c from the pending list}%
		\label{fig:bas2}
	\end{framed}
\vspace{-5mm}
\end{figure}

\subsubsection{Garbage collection and de-duplication of batch attestation shares} To limit the pending list from growing indefinitely and to avoid creation of two block headers corresponding to the same batch, the consensus nodes hold an in-memory database of digests of batches for which the number of corresponding batch attestations exceeded the $F+1$ threshold to create block headers. In other words, every time $F+1$ batch attestation shares are collected, the corresponding digest is added to the in-memory database. Then, consensus nodes know for which digests they should not attempt to create a second block header.
There are three problems to address with this design: (1) How to garbage collect the in-memory database to prevent it from growing indefinitely; (2) How do prevent malicious nodes from re-submitting old batch attestation shares after they have been garbage collected from the in-memory database, thus forcing creation of prior block headers in the hash chain; and lastly (3) how to uniformly and deterministically purge orphaned batch attestation shares from the pending list, since it is changed via the BFT consensus. 

To that end, Arma divides the time axis into discrete sections of equal length called epochs. Each section of time is represented by its epoch which is a monotonously ascending number.
When creating a batch attestation share, a batcher computes the current epoch and includes it in the payload of the batch attestation share sent to the consensus nodes.
Consensus nodes prevent totally ordering batch attestation shares with epochs too far in the past, therefore the second problem of re-submission of old batch attestation shares is avoided.
Arma addresses the first and third problems without relying on an assumption that the time in all correct consensus nodes is synchronized. Instead, old batch attestation shares are pruned via vote counting: Each batch attestation share of a shard contains references to orphaned batch attestation shares of the same shard as depicted in Figure \ref{fig:bas2}. Once a batch attestation share gathers $F+1$ votes that point to it, it is pruned from the pending list. A batch attestation share can only point to a batch attestation share of an earlier sequence number or of an earlier term, so cycles are avoided.

It is important to note that if the network is partitioned for a too long period of time, correct batchers may observe a batch attestation share in the pending list for which $F+1$ batch attestation shares have never been collected in the past, and as a result, falsely classify it as an orphaned batch attestation share, thus voting to purge it. 
In such a scenario, $F+1$ batch attestation shares may never be collected for the corresponding batch. A trivial way of addressing this, is having batchers that see their batch attestation shares being purged, re-submit them with new epochs. 

\begin{figure}[h]
	\fbox{
		\vspace{0pt}
		\begin{algorithm}[H] \label{alg:assembler}
			\SetKwBlock{DoParallel}{do in parallel}{end}
			\SetKw{Continue}{continue}
			\KwIn{\\$~~~~~$Streams of transaction batches from batcher nodes $\{\mathcal{B}_1, ..., \mathcal{B}_k\}$, one for each of the $k$ shards.\\
				$~~~~~$A stream of block headers from consensus nodes $\mathcal{H}$.\\
				$~~~~~$An index for blobs of data $\mathcal{I}$. Retrieves blocks by their digest $\mathcal{I}.Retrieve()$ in case they were indexed earlier by $\mathcal{I}.Index()$\\
				$~~~~~$A ledger of blocks $\mathcal{L}$.
			}	
			{
				
				\DoParallel{
					\ForEach{$shard\in \{1, .., k\}$}{
						$B \leftarrow \mathcal{B}_{shard}$ \; 
						$\mathcal{I}.Index(B)$ 
					}
					\For{true}{
						$h \leftarrow \mathcal{H}$\;
						\uIf{$\mathcal{I}.Exists(h.Digest)$}{
							$B \leftarrow \mathcal{I}.Retrieve(h.Digest)$\;
							$\mathcal{L}.Append(\big< h,  B\big>)$\;
						}\Else{
							\Continue	
						}
					}
				}				
			}	\label{fig:assembler}
			\caption{$~~$Pseudocode for assembling blocks from batch attestations and transaction batches}
		\end{algorithm}
	}	
\end{figure}

\subsection{Block assembly out of transaction batches and block headers produced by consensus}
\label{sec:assemb}
After each round of the Byzantine Fault Tolerant (BFT) consensus in
Arma, one or more block headers are persistently stored by the consensus nodes. These block headers include hashes referencing corresponding transaction batches, and monotonously ascending sequence numbers. The assembler nodes retrieve the block headers from the consensus nodes, while the transaction batches are retrieved from the batcher nodes.
Each block header carries a quorum of signatures from the consensus nodes, which the assembler nodes verify. 
This fact guarantees that all assembler nodes commit the same headers in the same order.
For each block header, an assembler node is responsible for retrieving
the corresponding batch, connecting them together to form a complete
block and subsequently writing it to the ledger as depicted in Algorithm \ref{fig:assembler}. 

By storing the entire block, which encompasses the block header, the associated
transactions, and a quorum of signatures on the block header (which contains a collision-resistant digest of the transactions), in the ledger, the assembler node is effectively an archivist of the Arma system.

As the shards independently generate batches at a potentially different
rate than the corresponding headers are totally ordered, it is impossible
to predict the exact number of batches fetched from a batcher relative
to the corresponding headers within a given time frame. Consequently,
it is not feasible to keep the batches in memory until their corresponding
block headers are retrieved. Thus, batches are promptly written to
disk once they are fetched. Similarly, there may be cases where
a block header is retrieved, but the corresponding batch is yet to
be retrieved. To address this, the block headers are also written
to disk instead of being stored in memory until the corresponding
batches are retrieved. Consequently, blocks are assembled
by reading the headers and batches from the disk rather than fetching
them from the network.

\subsection{Dynamic reconfiguration and onboarding}

The Arma system supports dynamic addition and removal of parties and
their corresponding nodes from the system by reconfiguration transactions
that can be submitted into the system. We consider the governance and the access control of these reconfiguration transactions to be out of scope, as this work focuses on the Arma protocol. 
Addition or removal of every node depends on its role:

\subsubsection{Node addition}
Each party can have as many assembler and router nodes as it likes, but only a single consensus node, as well as a single batcher per shard.
A new party is onboarded into Arma in the following manner:
\begin{enumerate}
	\item Its assembler node fetches the batch history and total order consensus
	history from assembler nodes of other parties.
	\item Once the assembler node catches up with other parties, the shards
	are expanded with batcher nodes from the joining party, all acting
	as secondary nodes to minimize the impact on the system. Unlike assembler
	nodes, batcher nodes need not replicate the entire batch history,
	therefore their addition is rather quick.
	\item Finally, the membership of the consensus set that run the Byzantine
	Fault Tolerant consensus protocol is expanded, adding a new consensus node.
\end{enumerate}
If the system is configured to withstand up to $F$ parties being
faulty, and the total number of parties is $3F+3$, then an addition
of party implicitly means the system can now withstand up to $F+1$
parties being faulty.

\subsubsection{Node removal}

Assembler nodes may be removed freely as long as they are not the
last assembler nodes removed from their own party, otherwise the party loses access to its own
source of newly created blocks.
If a consensus node is removed from the set of nodes that run the Byzantine Fault
Tolerant consensus, or a batcher node is removed from a shard, the
system can still operate as if the node has crashed. Removal of a
router node of a party impacts all its batchers, and is effectively
equivalent to having the entire party unreachable. Removal of a party
entirely involves removal of all of its nodes from the configuration
of the system. If the system is configured to withstand up to $F$
parties being faulty, and has exactly $3F+1$ parties taking part
in its execution, a removal of a party implicitly means that the system
can now only withstand $F-1$ parties being faulty. 

\subsubsection{Processing a reconfiguration transaction}

When a client submits a transaction to the router nodes of all parties,
the router nodes forward it to the appropriate batcher node for processing.
However, transactions that involve reconfiguring the system, such
as adding or removing nodes, should not be processed in the same manner
as ordinary transactions within the batcher shards. This distinction
arises due to a critical factor: If a reconfiguration transaction
is processed within a shard, the consensus nodes will only receive
the digest of the transaction and will not have access to the full
details necessary for processing the configuration change that may
directly impact them.

To address this issue, when a router node detects a reconfiguration
transaction, it redirects the transaction to the consensus node of
its respective party for processing, rather than routing it to the
batchers. This ensures that reconfiguration transactions are appropriately
handled and processed by the consensus nodes, allowing them to access
the complete information required for implementing the configuration
change.

\section{Integration into Hyperledger Fabric and performance evaluation}

In this section, we delve into how Arma can be integrated into Hyperledger Fabric. We then evaluate the performance of a prototype of a Hyperledger Fabric \cite{HLF} ordering service node embedding Arma components.

As previously mentioned, the Arma system comprises different
types of nodes, including router nodes, assembler nodes, batcher nodes,
and consensus nodes.
Conversely, in Hyperledger Fabric, there are only two types of nodes:
Peers, responsible for processing transactions, and ordering service nodes, tasked
with receiving transactions from clients and totally ordering
them into blocks for retrieval by the peers. Within this structure,
it is feasible to embed many of the Arma components within a Fabric
ordering node. While this topology may not fully leverage the scalability
capabilities of Arma, it still enables a substantial increase in the
throughput of the ordering service by an order of magnitude.

By integrating Arma into the Hyperledger Fabric framework and deploying
it within the ordering node, the ordering service can benefit from
the improved efficiency and performance offered by Arma's consensus
protocol. This integration represents a significant enhancement for
the ordering service in Hyperledger Fabric, allowing for a substantial
boost in transaction throughput.

In Hyperledger Fabric, ordering nodes perform two main tasks:

\begin{itemize}
	\item Verify transactions are well formed and properly signed by authorized clients.
	\item Totally order the transactions and bundle them into blocks signed by the orderer nodes themselves.
\end{itemize}

\begin{figure}[h]
	\begin{framed}
		\centering
		\includegraphics[width=14cm]{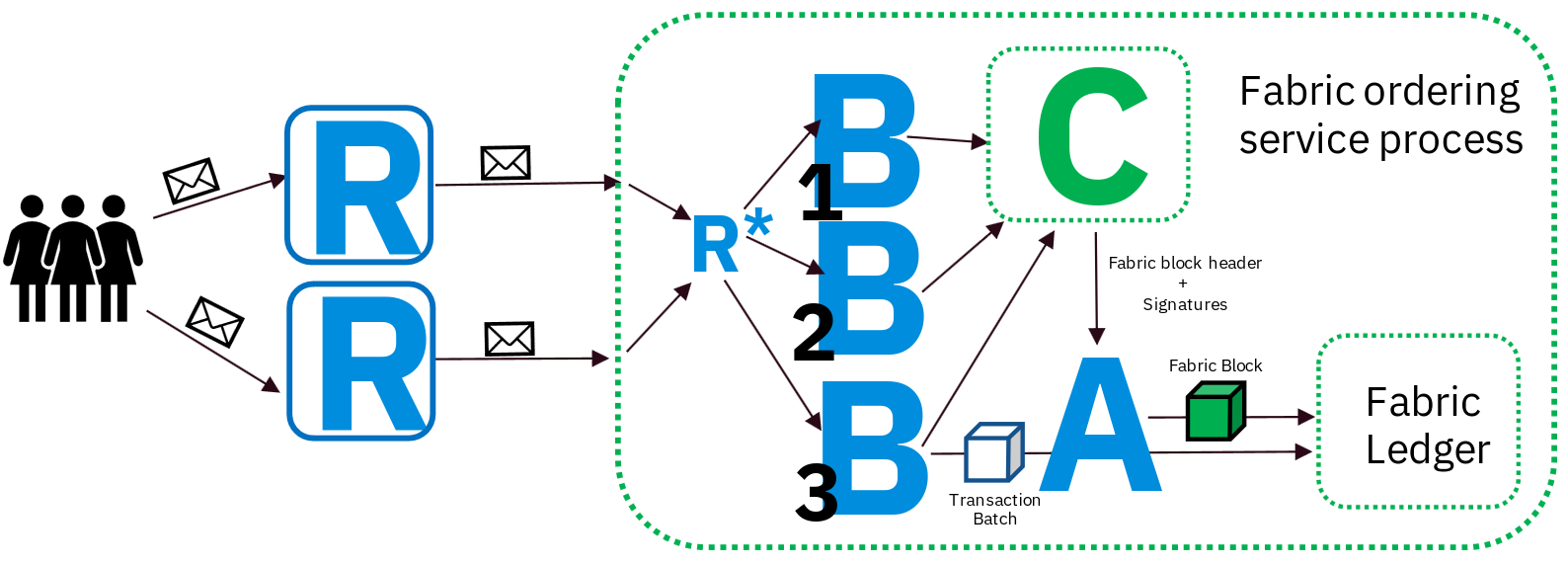}
		\caption{Arma components blue integrated alongside Fabric components (green) in a Fabric ordering node process.
			Transactions from clients are verified by router nodes and are forwarded into the ordering service node, where the $R^*$ router dispatches them to the batcher instances of appropriate shards.}%
		\label{fig:farma}
	\end{framed}
\end{figure}

We decouple these two tasks by delegating the former task of verifying transactions to Arma router nodes. 
The rest of the Arma nodes are embedded in the Hyperledger Fabric ordering service node, as seen in Figure \ref{fig:farma} .

As transactions arrive from the router nodes into the ordering node, they are routed to the batcher instance corresponding to the shard computed by the $R^*$ router logic. Then, transaction batches are written into the Fabric ledger as they are received from the primary batcher node. For the consensus node, we re-use Fabric's native BFT library \cite{SmartBFT} as-is, but modify its configuration to to order batch attestation shares instead of Fabric transactions and to produce batches of signed Fabric block headers. More specifically, in the original Hyperledger Fabric implementation, the BFT orderer embeds the SmartBFT \cite{SmartBFT} consensus library. The leader broadcasts an unsigned block and the nodes sign the block header and piggyback their signatures during the agreement protocol. As a result, a quorum of Fabric signatures is collected by every node at the end of an agreement on a block. In our prototype, we made the SmartBFT leader node broadcast a batch of batch attestation shares (and complaints on primary batcher nodes), and instead of signing over the entire batch, they deterministically assemble Fabric block headers and sign over them. The resulting block headers are then passed into the assembler component. Afterwards, the transaction batches are retrieved from the Fabric ledger, and full Fabric blocks are assembled and written to the ledger again, ready for retrieval by peers.

\subsection{Chaining transaction batches in Hyperledger Fabric}
In some distributed ledgers it is not possible to associate a block header with more than one batch. For example, in Hyperledger Fabric
each block header contains essential information such as a sequence
number, a hash of the previous header, and a hash of the transactions
associated with the block. The transactions are concatenated and then
hashed using a cryptographic hash function. Due to the structure of
a block header, which consists of a single hash, it is not possible
to associate a Fabric block header with more than one batch. Hence,
When integrated with Hyperledger Fabric, Arma's consensus nodes generate a single Fabric block header for each
$F+1$ batch attestation shares.
Within each set of block headers that is totally ordered via the Byzantine
Fault Tolerant consensus protocol, the Fabric block headers are linked
together in a hash chain.

\subsection{Performance evaluation}
We evaluate the performance of an Arma prototype integrated into a Hyperledger Fabric ordering service node. We have incorporated batchers, assemblers, and consensus nodes directly into the ordering service node process. Additionally, we have represented a router node as a straightforward function that maps transactions to the appropriate batcher instance within the Fabric ordering service node process.

It's important to note that in a real-world production environment,  the router component would ideally operate on a dedicated node and perform transaction verifications as transactions arrive from clients. Our performance evaluation primarily focuses on measuring the efficiency with which transactions originating from clients are totally ordered into Fabric blocks, without the involvement of router nodes in the verification process.

It's worth mentioning that while our evaluation involved embedding all components within a single ordering service node for simplicity, this may not be the most scalable and performant deployment strategy for Arma. We recognize the need for future work to assess the performance of a distributed deployment of Arma where each node role runs in its own machine.

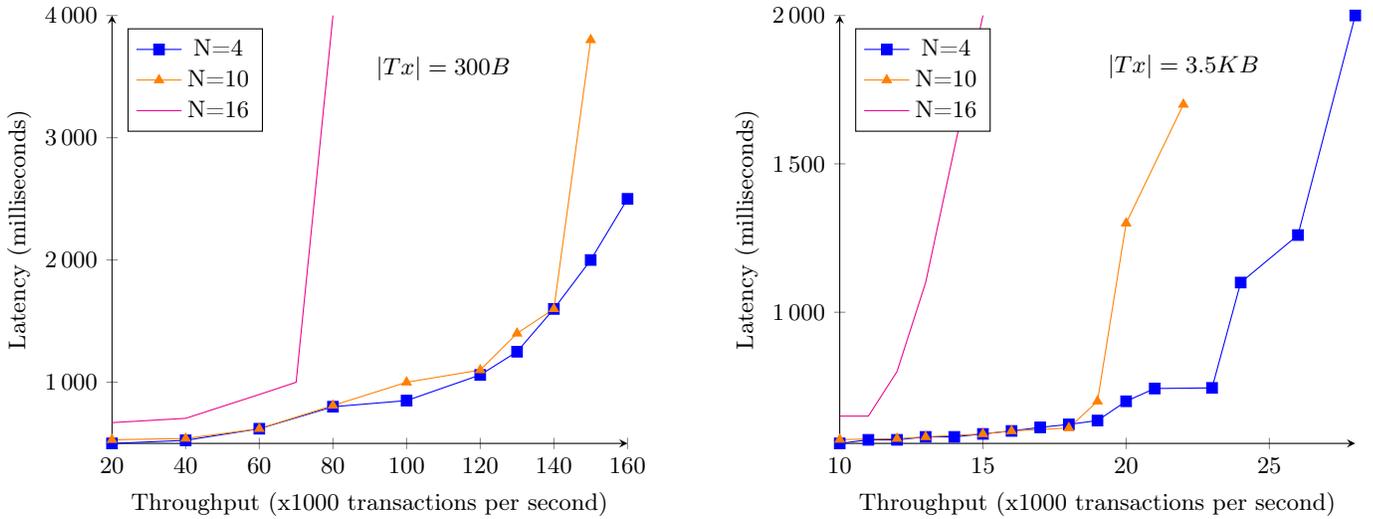
\begin{figure}[h]
	\centering
	\begin{minipage}[t]{.45\linewidth}
		\begin{tikzpicture}
			\begin{axis}[
				log ticks with fixed point,
				legend pos = north west,
				axis lines = left,
				xlabel = Throughput (x1000 transactions per second),
				ylabel = Latency~(milliseconds),
				/pgf/number format/1000 sep={\,},
				]
				
				\node[above] at (90,290) {$|Tx|=300B$};

				\addplot [
				color=blue,
				mark=square*,
				]
				coordinates {
					(20, 500)(40, 525)(60, 620)(80, 800)(100, 850)(120, 1060)(130, 1250)(140, 1600)(150, 2000)(160, 2500) 
				};
				\addlegendentry{N=4}				
				
				\addplot [
				color=orange,
				mark=triangle*,
				]
				coordinates {
					(20, 530)(40, 540)(60, 620)(80, 810)(100, 1000)(120, 1100)(130, 1400)(140, 1600)(150, 3800)
				};
				\addlegendentry{N=10}			
				
				\addplot [
				color=magenta,
				mark=circle*,
				]
				coordinates {
					(20, 670)(40, 705)(60, 900)(70, 1000)(80, 4000)
				};
				\addlegendentry{N=16}

			\end{axis}
		\end{tikzpicture}
	\end{minipage}
	\hfill
	\begin{minipage}[t]{.45\linewidth}
		\begin{tikzpicture}
			\begin{axis}[
				log ticks with fixed point,
				legend pos = north west,
				axis lines = left,
				xlabel = Throughput (x1000 transactions per second),
				ylabel = Latency~(milliseconds),
				/pgf/number format/1000 sep={\,},
				]
				
				\node[above] at (120,120) {$|Tx|=3.5KB$};
				
				\addplot [
				color=blue,
				mark=square*,
				]
				coordinates {
					(10, 558)(11, 570)(12, 570)(13, 580)(14, 580)(15, 590)(16, 600)(17, 612)(18, 622)(19, 635)(20, 700)(21, 743)(23, 745)(24, 1100)(26, 1260)(28, 2000)
				};
				\addlegendentry{N=4}

				\addplot [
				color=orange,
				mark=triangle*,
				]
				coordinates {
					(10, 571)(12, 574)(13, 580)(15, 590)(16, 600)(18, 610)(19, 700)(20, 1300)(22, 1700)
				};
				\addlegendentry{N=10}

				\addplot [
				color=magenta,
				mark=circle*,
				]
				coordinates {
					(10, 650)(11, 650)(12, 800)(13, 1100)(15, 2000)
				};
				\addlegendentry{N=16}	
				
			\end{axis}
		\end{tikzpicture}
	\end{minipage}
	\caption{Performance evaluations with varying number of nodes ($N$) and transactions varying in size}
	\label{fig:evaluation}						
\end{figure}

\subsubsection{Experiment setup}
The evaluation was carried out using multiple clients situated in England, generating transactions of two different sizes: 300 bytes in one experiment and 3.5 KB, which corresponds to the standard Hyperledger Fabric transaction size, in another. Each transaction was sent to all ordering service nodes in parallel. Subsequently, the clients retrieved blocks from these ordering nodes and calculated both throughput and latency.

The ordering service nodes were deployed across three distinct datacenters located in England, Italy, and France. Latency measurements between these datacenters were as follows: 10ms between England and France, 20ms between England and Italy, and 17ms between Italy and France.

For the infrastructure, the ordering service nodes were hosted on dedicated bare-metal Ubuntu 22.04 LTS machines. These machines were equipped with 96 Intel Xeon 8260 2.40GHz processors (comprising 48 cores with 2 threads per core) and 64GB of RAM. Additionally, they featured a Broadcom 9460-16i RAID 0 configuration with two SSDs.

We conducted experiments with various numbers of Arma ordering service nodes starting from 4 nodes up until and including 16 nodes.

\subsubsection{Result Analysis}
The results of the evaluation can be seen in Figure \ref{fig:evaluation} . The right and left graphs show evaluation with transaction sizes of 300 bytes and 3500 bytes respectively. 

First, it can be clearly seen that the higher the amount of nodes the lower the throughput. This is expected, as more data needs to be transmitted among the nodes.

An interesting fact is while in the experiment with the 3.5KB transaction size, the transactions are $11.6$ times bigger than the experiment with the 300B transaction size, the difference in transactions per second totally ordered is only $\sim$ 6 times higher for the smaller transactions. This may imply an overhead that stems from transaction batching and censorship resistance mechanisms, since the amount of data totally ordered in the experiment of the large transactions is twice bigger than the other experiment.
However, it could be associated to the fact that parameters such as batch size and maximum batch latency in both experiments were identical, and calls for further fine tuning of parameters as a function of the transaction size.

\section{Conclusions and future work}

At its essence, Arma formulates a technique for amplifying the performance of a consensus protocol by decoupling data dissemination from the actual consensus mechanism. As demonstrated in the evaluation of the prototype, it exhibits considerable potential in enhancing the scalability and performance of consensus. Two key insights that stem from this research are in the ability to enhance the performance of a slow, non-pipelined consensus like SmartBFT \cite{SmartBFT} by employing the methodologies explained in Arma, and how to add censorship resistance to an Arma like system.
However, it is important to note that the current iteration of the prototype falls short in certain aspects, potentially influencing the evaluation outcomes. A prominent issue is that all Arma components run by a party were integrated within the same Fabric node. Subsequent iterations of this research will evaluate a fully distributed Arma deployment, which holds promise for further performance enhancements, as pointed out by the findings of 'Narwhal and Tusk' \cite{Narwhale}. Additionally, proper input sanitation would be implemented in order to ascertain whether it has an effect on the performance. 

Lastly, a comparative analysis between a purely distributed Arma deployment and a Fabric-compatible Arma deployment is needed to quantify the extent to which the Fabric stack influences performance.

\bibliographystyle{unsrt}
\bibliography{references}

\end{document}